\documentclass[12pt,aps,preprint,showpacs]{revtex4}

\usepackage{dcolumn}

\usepackage{graphicx}

\vspace*{1.5cm}

\usepackage[centertags]{amsmath}
\usepackage{amssymb}

\begin{document}

\title{Probing the Gravitational Scale via Running Gauge Couplings}

\author{Ilia Gogoladze\footnote{On a leave of absence from:
Andronikashvili Institute of Physics, GAS, 380077 Tbilisi, Georgia.}}
\email{ilia@physics.udel.edu} \affiliation{\it Department of Physics
and Astronomy, University of Delaware, Newark, DE 19716, USA}

\author{C.N. Leung}
\email{leung@physics.udel.edu} \affiliation{\it Department of
Physics and Astronomy, University of Delaware, Newark, DE 19716, USA}

\date{\today}

\begin{abstract}

According to a recent paper by Robinson and Wilczek, the leading
gravitational corrections to the running of gauge couplings tend
to reduce the values of the couplings at energies below the
gravitational scale, defined to be the energy above which gravity
becomes strongly interacting.  If the physical gravitational scale
is sufficiently low, as conjectured in certain extra-dimension
models, this behavior of the gauge couplings can be measured in
future high energy experiments, providing a way to determine
where the gravitational scale lies.  We estimate that measurements
of the fine structure constant at the Large Hadron Collider and
the proposed International Linear Collider energies can probe
the gravitational scale up to several hundred TeV, which will be
the most stringent test that can be obtained in the conceivable
future.

\end{abstract}


\maketitle

In a recent paper~\cite{RW}, Robinson and Wilczek have examined the
modification to the running of gauge couplings due to gravitational
corrections~\cite{Odintsov:1994rb}.  At the one-loop level, they
found that the effective gravitational corrections amount to a
negative term in the beta function of the gauge coupling $g$:
\begin{equation}
 \beta(g, E) \equiv \frac{dg}{d \ln E} = - \frac{b_0}{(4 \pi)^2} g^3
 - \frac{3}{\pi} \left(\frac{E}{\Lambda_G}\right)^2 g,
\label{betafcn}
\end{equation}
where $E$ is the energy, $\Lambda_G$ is the gravitational scale, the
energy at which gravitational interactions become strong, and $b_0$
is the usual coefficient of the $\beta$ function in the absence of
gravity.  The gravitational correction is the same for all gauge
couplings because gravitons carry no gauge charge.  If one assumes
that in a consistent quantum theory of gravitational interactions,
which is yet to be discovered, the above one-loop contributions
represent the leading corrections, then all gauge couplings will
be driven toward zero at energies around $\Lambda_G$.  Thus,
gravitational interactions have the tendency to asymptotically
unify all gauge couplings.  While this perturbative result is not
reliable at $\Lambda_G$, it should be a reasonable approximation at
energies sufficiently below $\Lambda_G$.  And if this negative
running of the gauge couplings causes sizable corrections at energies
sufficiently below $\Lambda_G$, it may be detectable in future high
energy experiments, depending on what the value of $\Lambda_G$ is.

While it is usually assumed that $\Lambda_G$ is of order the Planck
energy, $M_P \equiv \sqrt{\hbar c/G_N} = 1.2 \times 10^{19}$~GeV,
where $\hbar$ is the Planck constant divided by $2 \pi$, $c$ is the
speed of light in vacuum, and $G_N$ is the Newton gravitational
constant, one cannot be certain until a consistent quantum theory
of gravity is established.  There are proposals that involve the
assumed existence of extra spatial dimensions~\cite{ADD}, in which
the physical gravitational scale may be much lower than $M_P$.
For instance, with $n$ extra dimensions compactified on a torus of
radius $R$, the gravitational scale is given by~\cite{ADD}
\begin{equation}
\Lambda_{\small (ADD)}^{n+2} =\frac{\hbar c}{G_N}
\left(\frac{\hbar}{c}\right)^n \frac{1}{(2 \pi R)^n}
\end{equation}
where $(2 \pi R)^n$ is the volume of the n-torus.  Thus, the
gravitational scale can be sufficiently low if $R$ and/or $n$ is
sufficiently large.  The current experimental upper bound on $R$
is $0.13$~mm~\cite{limit}, for $n = 2$.  This corresponds to a
lower bound on $\Lambda_G$ of order 1~TeV.

If $\Lambda_G$ is indeed many orders of magnitude smaller than
$M_P$, the gravitational effects on the running of gauge couplings
may be measurable in future collider experiments and be used to
determine the value of $\Lambda_G$.  This is what we will examine
below.  What we are primarily interested in is to examine the
feasibility of determining $\Lambda_G$ by this method and to
estimate the potential reach of the Large Hadron Collider (LHC),
rather than to test specific models.  We will therefore use the
result~(\ref{betafcn}) as our testing ground and neglect
additional terms that may arise in specific models (e.g., there
may be additional corrections from the KK gravitons in
extra-dimension models).  In other words, we use the result
(\ref{betafcn}), which was obtained from Einstein gravity, as
a generic description of the running of the gauge couplings
even though it may not be an accurate description of the running
in a specific model.  And, instead of assuming $\Lambda_G$
to be the Planck energy, as was done in Ref.~1, we let it be a
free parameter and examine what limits on $\Lambda_G$ may be
accessible from precision measurements of the gauge couplings in
future colliders such as the LHC.

\begin{figure}[htb]
\centering
\includegraphics[width=16cm]{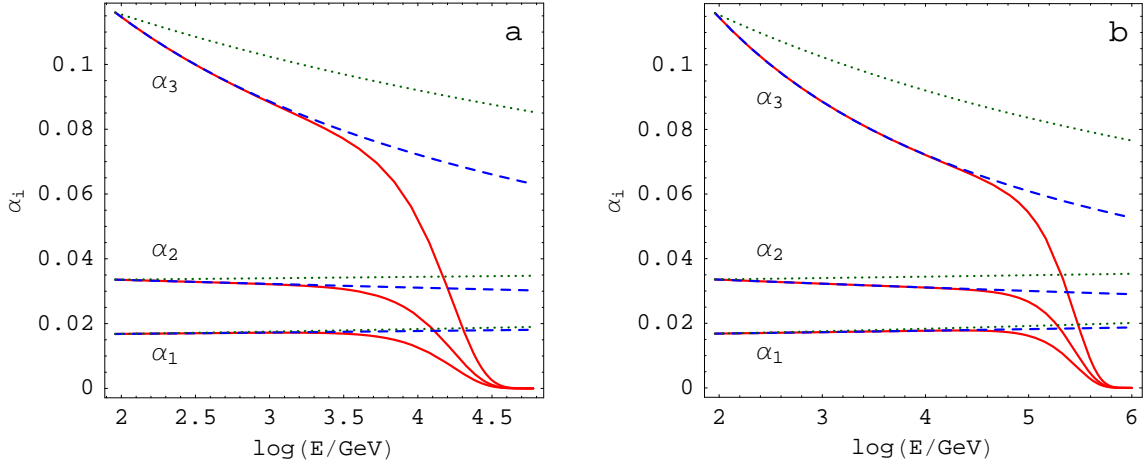}
\caption{Running of Standard Model gauge couplings with (solid
curves) and without (dashed curves) gravitational corrections,
for (a) $\Lambda_G = 35$~TeV and (b) $\Lambda_G = 10^3$~TeV.
For comparison, the running of the gauge couplings in the
minimal supersymmetric standard model is also shown (dotted
curves).}
\label{smGmssm}
\end{figure}

Figure~\ref{smGmssm} compares the running of the gauge couplings
in the Standard Model with and without the gravitational
corrections for various values of $\Lambda_G$.  Here $\alpha_i =
g_i^2/(4 \pi)~(i = 1, 2, 3)$, where $g_1$, $g_2$, and $g_3$ are
the couplings of the gauge groups $U(1)_Y$, $SU(2)_L$, and
$SU(3)_C$, respectively.  The evolution of the couplings are
obtained by integrating Eq.~(\ref{betafcn}) from the initial
energy of $M_Z$, the mass of the $Z$ boson, with the initial
data~\cite{Eidelman:2004wy}:
\begin{eqnarray}
&&\alpha^{-1}(M_Z) = 128.91 \pm 0.02\,, \nonumber \\
&&\sin^2\theta_W(M_Z) = 0.23120 \pm 0.00015\,, \nonumber \\
&&\alpha_3(M_Z) = 0.1182 \pm 0.0027\,,
\label{input}
\end{eqnarray}
where $\alpha$ is the fine structure constant and $\theta_W$ is
the weak mixing angle.  Since we are only interested in an
estimate of the size of the one-loop gravitational corrections,
we take only the one-loop contributions to the coefficient $b_0$
in Eq.~(\ref{betafcn}), with $b_0 = 7$ for $SU(3)_C$, $b_0 = 19/6$
for $SU(2)_L$, and $b_0 = -41/6$ for $U(1)_Y$, assuming three
generations of quarks and leptons.  In all cases, we see that
the gravitational corrections become significant at energies
which are more than one order of magnitude below $\Lambda_G$,
energies for which the effective gravitational corrections
from~(\ref{betafcn}) may be considered reliable.  This behavior
persists for values of $\Lambda_G$ larger than those illustrated
in Figure~\ref{smGmssm}.

\begin{figure}[htb]
\centering
\includegraphics[width=16cm]{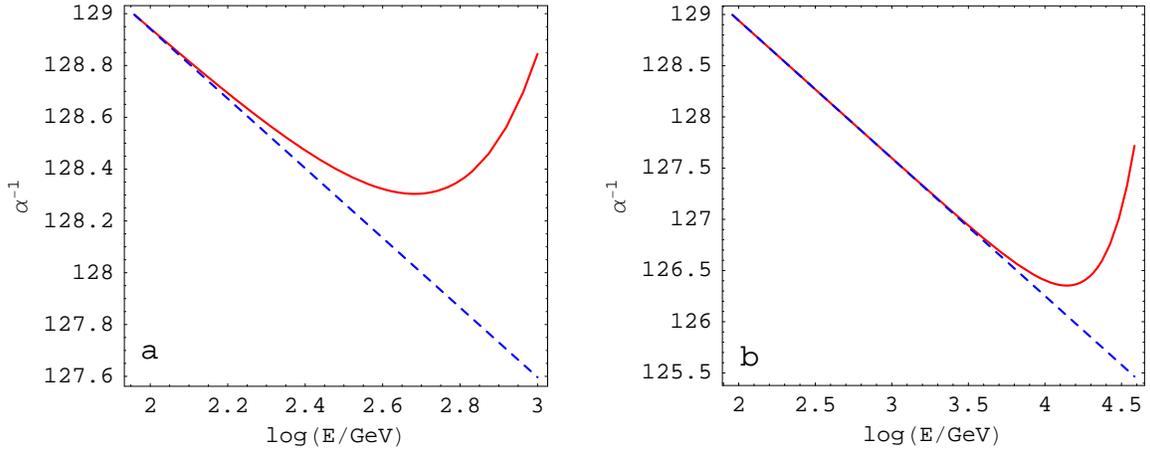}
\caption{Running of the fine structure constant with (solid)
and without (dashed) gravitational corrections, for (a)
$\Lambda_G = 35$~TeV and (b) $\Lambda_G = 10^3$~TeV.}
\label{alpha}
\end{figure}

\begin{figure}[htb]
\centering
\includegraphics[width=16cm]{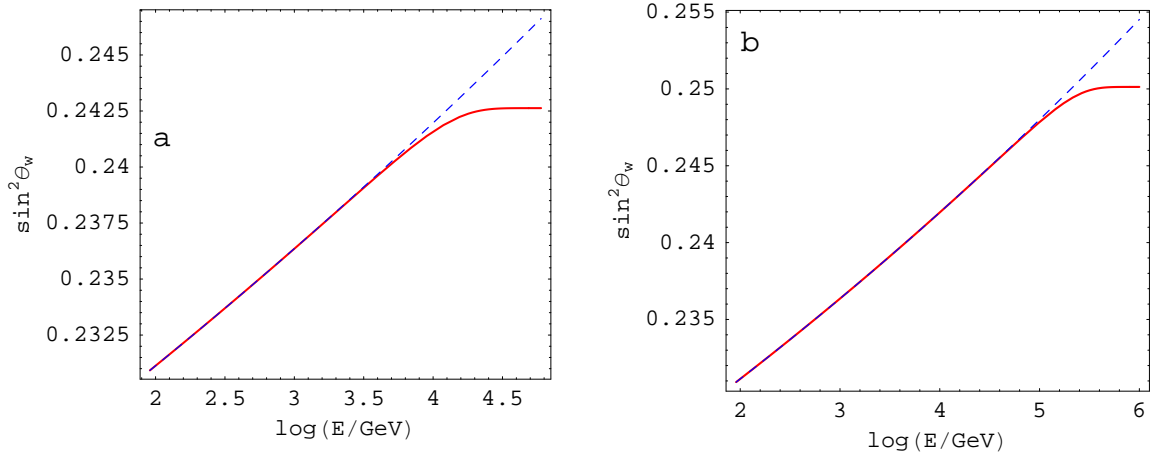}
\caption{Running of $\sin^2\theta_W$ of the Standard Model
with (solid) and without (dashed) gravitational corrections,
for (a) $\Lambda_G = 35$~TeV and (b) $\Lambda_G = 10^3$~TeV.}
\label{sinW}
\end{figure}

Instead of $\alpha_1$ and $\alpha_2$, it may be more useful
to examine $\alpha$ and $\sin^2 \theta_W$, which have been
measured to high precisions.  The running of these two
quantities are shown in Figures~\ref{alpha}~and~\ref{sinW}.
Because $\alpha$ is more precisely known, and hence can
provide the more stringent constraint, it is an ideal
probe for the gravitational scale.  As an illustration,
if we require that the fractional change in $\alpha^{-1}$
due to the gravitational corrections at $E = 200$~GeV
(see Table 1 below) cannot exceed twice the fractional
error obtained from~(\ref{input}), we find that $\Lambda_G$
must be at least 35~TeV.  If we assume the same error as
in~(\ref{input}) will be achieved in measuring $\alpha$
at $E = 200$~GeV, the lower bound on $\Lambda_G$ will be
raised to about 60~TeV.

\renewcommand{\arraystretch}{1.4}
\begin{table}[ht]
\vspace{0.4cm}
\begin{center}
\begin{tabular}{|c|c|c|c|}
\hline E &~~ $100 \times
\frac{\vert\Delta\alpha^{-1}\vert}{\alpha^{-1}}$~~ &~~ $100\times
\frac{\vert\Delta \sin^2\theta_W \vert}{\sin^2\theta_W }$~~
&~~ $100 \times \frac{\vert\Delta\alpha_3\vert}{\alpha_3}$~~\\
\hline
2 TeV & 3.997& 0.03949& 3.662\\
\hline
1 TeV & 0.978 &0.00969 &0.921\\
\hline
 500 GeV & 0.237 &0.00219 &0.225\\
\hline
 ~~200 GeV~~ &0.031  & 0.00019 &0.030\\
\hline
\end{tabular}
\end{center}
\vspace{-0.3cm} \caption{\small Gravitational corrections to
the running of gauge couplings for $\Lambda_G = 35$~TeV.
\label{Spectrum-UP-7D}}
\end{table}

For $\Lambda_G = 35$~TeV, we show in Table 1 the fractional
changes in $\alpha^{-1}$, $\sin^2 \theta_W$, and $\alpha_3$
arising from the gravitational corrections at various energies.
For each quantity $x$, the displayed $\frac{\vert \Delta x
\vert}{x} \equiv \vert (x$~with gravitational
corrections~$-~x$~without gravitational corrections)$\vert$
divided by $x$ without gravitational corrections.  Because the
changes increase with energies, future high energy experiments at
the LHC and beyond, e.g., the proposed International Linear
Collider (ILC), will be able to probe larger values of
$\Lambda_G$, depending on the precisions that can be achieved in
measuring these quantities.  For example, assuming the error in
measuring $\alpha^{-1}$ at $1.5$~TeV, which is a representative
energy for the LHC and the ILC, will be twice the error given
in~(\ref{input}), we find that $\Lambda_G$ must be at least
300~TeV in order that the gravitational correction does not exceed
this assumed experimental error.  Obviously, a smaller $\Lambda_G$
will be allowed by a larger experimantal error.  As an
illustration, if the error in measuring $\alpha^{-1}$ at $1.5$~TeV
were five times the current error in~(\ref{input}), the lower
bound on $\Lambda_G$ would become 200~TeV.

In comparison, the range of $\Lambda_G$ that can be probed
by the running coupling of the strong interaction will be
lower.  For instance, assuming the error in measuring
$\alpha_3$ at $1.5$~TeV will be two (five) times the error
given in~(\ref{input}), we find that $\Lambda_G$ must be at
least 27~TeV (18~TeV) in order that the gravitational
correction does not exceed this assumed error.

We highlight the very important feature in the numerical
examples above that limits on $\Lambda_G$ are derived from
the corresponding gravitational corrections at energies
which are at least two orders of magnitude below the cutoff
scale $\Lambda_G$, thus making them reliable predictions.

We have also considered the case of the minimal
supersymmetric standard model (MSSM), for which the
running of the gauge couplings differs from the Standard
Model.  We assume for simplicity that the {\it effective}
supersymmetry breaking scale is $M_{Z}$~\cite{Langacker:1992rq}.
We compare in Figure~\ref{smGmssm} the MSSM running with the
gravitationally corrected Standard Model running.  The two
cases can clearly be distinguished.  In particular, the
gravitational corrections affect all gauge couplings the same
way, which is not the case for MSSM.  There may be other
extensions of the Standard Model that would affect the
running of the gauge couplings.  However, it is unlikely any
of these extensions will share the same characteristics as
the gravitational correction, which is to reduce every gauge
coupling in a similar manner.  It will therefore be relatively
straightforward to identify the gravitational effects.

In conclusion, we have shown that the effective gravitational
corrections to the gauge coupling running can be detected at
energies significantly below the gravitational scale
$\Lambda_G$.  If $\Lambda_G$ is no more than several hundred
TeV, such effects may be measurable at the LHC or the ILC,
signaled by values of the gauge couplings which are smaller
than the Standard Model expectations.  It is very interesting
that a precise measurement of the running gauge couplings can
provide information about the nature of the gravitational
interactions and yield the most stringent limit on the scale
of quantum gravity attainable in the conceivable future. \\

%\newpage

This work was supported in part by the U.S. Department of
Energy under Grant DE-FG02-84ER40163.  We thank K.~S.~Babu,
B.-L.~Hu, S.~T.~Love, C. Macesanu, K.-W.~Ng, S.-P.~Ng and
Y.~J.~Ng for useful discussions and D. Chung for thoughtful
comments about our work.

\end{document}